\documentclass[10pt,aps,prl,twocolumn,superscriptaddress,nofootinbib,tightenlines,floatfix,titlepage]{revtex4-2}

\usepackage[T1]{fontenc}
\usepackage[utf8]{inputenc}
\usepackage{amsmath,amssymb,amsfonts,bm}
\usepackage{graphicx}
\usepackage{xcolor,ulem}
\usepackage{hyperref}
\usepackage{orcidlink}
\usepackage{booktabs}
\usepackage{multirow}
\usepackage{tikz}
\usetikzlibrary{positioning}
\usepackage{cleveref}
\usepackage{chngcntr}

\hypersetup{colorlinks=true,linkcolor=blue,citecolor=blue,urlcolor=blue}

\newcommand{\Sf}{S_{\rm f}}
\newcommand{\Sc}{S_{\rm c}}

\newcommand{\ee}{\mathrm{ee}}

\newcommand{\figplaceholder}[2]{%
\IfFileExists{#1}{\includegraphics[width=\linewidth]{#1}}{%
\fbox{\begin{minipage}[c][0.30\textheight][c]{0.92\linewidth}\centering
\texttt{#1}\\[0.5em]#2/
\end{minipage}}}}

\begin{document}

\title{Renormalization-guided cascade upscaling for lattice field generation}

\author{Anna Hasenfratz\,\orcidlink{0000-0003-1813-2645}}
\email[Corresponding author: ]{anna.hasenfratz@colorado.edu}
\affiliation{Department of Physics, University of Colorado Boulder,
Boulder, Colorado 80309, USA}

\author{Ethan T.~Neil\,\orcidlink{0000-0002-4915-3951}}
\email[Corresponding author: ]{ethan.neil@colorado.edu}
\affiliation{Department of Physics, University of Colorado Boulder,
Boulder, Colorado 80309, USA}

\author{Letizia Parato\,\orcidlink{0000-0001-7500-6747}}
\email{letizia.parato@colorado.edu}
\affiliation{Department of Physics, University of Colorado Boulder,
Boulder, Colorado 80309, USA}

\author{Noah Schwartz}
\email{noah.schwartz-2@colorado.edu}
\affiliation{Department of Physics, University of Colorado Boulder,
Boulder, Colorado 80309, USA}
\date{\today}

\begin{abstract}
We introduce a renormalization-group (RG) guided machine-learning algorithm for lattice field generation based on approximate inversion of an RG transformation.  A ``perfect blocking'' construction supplies equilibrated long-distance modes, while a conditional normalizing flow reconstructs short-distance details and brief rethermalization removes residual errors.  In 2D $\phi^4$ theory at criticality, a flow trained at $L\le32$ is reused recursively in cascades reaching $L=2048$ with correct long-distance physics.
\end{abstract}

\maketitle

\paragraph{Introduction --} Markov-chain generation of lattice field configurations becomes increasingly expensive near a critical point, a phenomenon known as critical slowing down \cite{Schaefer:2010hu,Finkenrath:2024ptc}.  This effect obstructs the numerical study of quantum critical points, which appear in a wide range of interesting condensed matter and particle physics systems, including in taking the continuum limit in lattice quantum chromodynamics (QCD).  
  The problem of critical slowing down is especially acute for long-distance observables, which evolve only diffusively under Markov-chain algorithms that apply local updates and suffer from autocorrelation times that scale rapidly with the system size.  
Solutions such as the Wolff cluster update \cite{Wolff:1989} can greatly mitigate critical slowing down in spin systems, but there is no known way to generalize such methods to gauge or fermion degrees of freedom.

\begin{figure*}[t]
    \centering
    \includegraphics[width=0.95\textwidth]{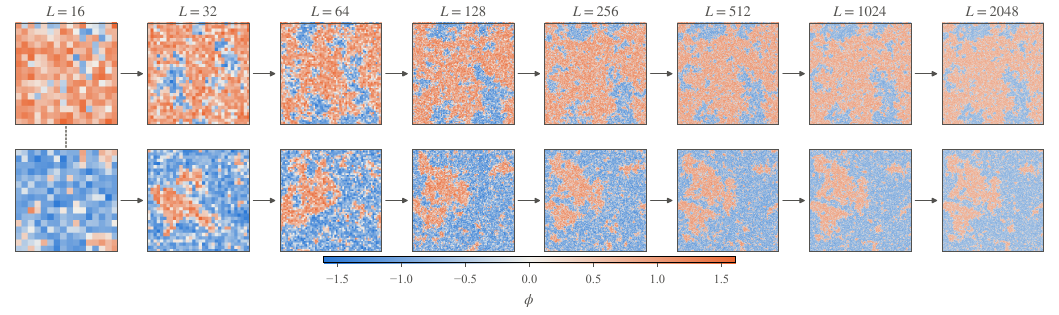} 
    \caption{Recursive inverse-blocking cascades in the two-dimensional $\phi^4$ model.
    Starting from equilibrated $L=16$ configurations at $\kappa_{\rm{cr}}$, each scale-two upscaling
    step is followed by HMC rethermalization targeting the fine action.
    Two independent cascades are shown; the long-distance structure can be seen developing from small to large volume.
    }
    \label{fig:upscale_tower}
\end{figure*}

Recently, normalizing-flow based sampling methods \cite{Albergo:2019eim,Albergo:2021vyo,DelDebbio:2021qwf,Komijani:2023fzy} have been investigated as an alternative approach, using machine learning to construct approximate samplers that draw configurations directly from the target Boltzmann distribution with density \(e^{-S(\phi)}\). These methods allow one to draw samples with minimal autocorrelation, and have been shown to generalize fully to gauge and fermion theories, including QCD \cite{Boyda:2020hsi,Albergo:2021bna,Abbott:2022zhs,Abbott:2022zsh}.  However, in practice, normalizing-flow sampling also faces severe problems with volume scaling.  Whole-volume flows near a critical point must encode fluctuations at all distance scales, and errors in this encoding must be corrected using a global accept/reject step.  This results in global acceptance which deteriorates exponentially with the system volume, unless the training is made increasingly accurate to compensate.  In effect, critical slowing down remains in this approach but is pushed into the flow training problem.

In this work, we approach the volume-scaling problem in a different way by mapping coarse to fine lattice ensembles. Specifically, we construct the coarse ensemble to approximate the probability distribution obtained by renormalization-group (RG) blocking of the target fine theory. In this way, the coarse ensemble already encodes the correct long-distance degrees of freedom, and the subsequent upscaling procedure is required only to reconstruct the missing short-distance information. In image processing, increasing resolution by filling in such fine details is known as ``upscaling.'' In RG blocking, we separate short-distance averaging or smoothing from coarse graining or decimation and use a conditional normalizing flow to reconstruct the degrees of freedom removed by decimation, thereby implementing an approximate inverse RG transformation.

 Generally, RG blocking does not preserve the initial lattice action; instead, it flows the system in the infinite-dimensional space of couplings \footnote{In the language of continuum effective field theory, RG blocking integrates out modes above a cutoff $\Lambda$, inducing an infinite number of effective ``irrelevant'' operators suppressed by powers of $\Lambda$.}. 
This makes it difficult to generate coarse configurations with the distribution required for upscaling. One way to mitigate the problem is to learn the form of the more complicated fixed-point action directly, which is known as the ``perfect action'' approach~\cite{Hasenfratz:1993sp,DeGrand:1995jk};
this has also been explored recently using machine-learning methods in \cite{Holland:2024muu,Holland:2025fsa}.  We instead pursue the idea of ``perfect blocking'', adjusting the definition of the RG block transformation itself so that the blocked action remains as close as possible to the form of the initial lattice action.

Both the invertible smoothing kernel that defines the perfect blocking and the conditional normalizing flow act quasi-locally on the lattice fields, which allows them to generalize well from small volumes to larger ones.  In our proposed algorithm for generating lattice ensembles at criticality, which we call ``cascade upscaling'', a decorrelated small-volume ensemble with linear extent $L$, generated with traditional Monte Carlo methods, is repeatedly upscaled to give independently seeded large-volume configurations in the cascade $L \rightarrow 2L \rightarrow 4L \rightarrow 8L ...$, reaching system sizes far larger than the ensembles used for training the kernel and normalizing flow.  Short-distance imperfections in the  inverse RG blocking are corrected at each step using ordinary Monte Carlo algorithms to ``rethermalize'' according to the target fine action.

In addition to describing the general method, we demonstrate cascade upscaling in a numerical example, working in the two-dimensional $\phi^4$ theory tuned to the critical point.  The example field configurations in \cref{fig:upscale_tower} visually show the results of this process.  As can be seen, the resulting long-distance structure of the large-volume configurations develops from small to medium volumes where the local Monte Carlo rethermalization acts efficiently.  Moreover, each small-volume chain seeds an independent cascade, so that a cascade seeded from fully decorrelated small-volume configurations produces decorrelated large-volume samples.

Our strategy is related in spirit to multiscale equilibration and prolongation methods~\cite{Endres:2015yca} and to inverse-renormalization or super-resolution ideas~\cite{PhysRevLett.89.275701,Bachtis:2021eww,Marchand:2022fxp,Wang:2023exq,Bauer:2024byr,Zhu:2025pmw,Masuki:2025pfx,Tan:2026ibs,Masuki:2026qfw,Singha:2026slc}.  The crucial new ingredient is the combination of an (approximate) perfect blocking transformation, a conditional model for the missing detail fields, and correction to the target fine action using local Monte Carlo updates.  Each of the key required ingredients in our algorithm - RG blocking, normalizing flows, and rethermalization - all have natural extensions to gauge theories and, in principle, to fermionic systems. Moreover, the construction does not require exact tuning to criticality.

\paragraph{Basic definitions and blocking -- } For concreteness, we work with $\phi^4$ scalar field theory to describe our proposed algorithm.  The lattice action is 
\begin{equation}
  S[\phi]= -2\kappa\sum_{x,\mu}\phi_x\phi_{x+\hat\mu}
  +\sum_x \left[\phi_x^2+\lambda(\phi_x^2-1)^2\right] .
  \label{eq:phi4-action}
\end{equation}

An RG blocking transformation $B$ can be usefully divided into two steps: a smoothing kernel $K$ and a decimation $D_s$ which reduces the number of degrees of freedom by scale $s$; we fix $s=2$ for the remainder of this work, so that $D_2$ reduces the degrees of freedom as $D_2: L^d \rightarrow (L/2)^d$.  Given a field configuration $\phi$ of size $L^d$ the two operations act as
\begin{equation}
\psi = K \phi;\, \, \phi_b = D_2 \psi = D_2 K \phi
\end{equation}
so that $B \equiv D_2 K$.  Specializing to $d=2$ for simplicity, we can decompose the smoothed field $\psi$ as
\begin{equation}
    \phi
    \longleftrightarrow
    \psi = (\phi_b,d_{01},d_{10},d_{11}) = (\psi_{ee}, \psi_{eo}, \psi_{oe}, \psi_{oo}).
\end{equation}
where $e,o$ subscripts denote even or odd coordinates in each direction. The blocked field $\phi_b$ is selected to be the even-even sites of $\psi$, while we refer to the other fields $d_{ab}$ as the ``fine-detail fields''.  We note that the detail fields are similar to $\phi_b$ and could also have been used as candidate blocked fields; the ultraviolet degrees of freedom lie in the differences $d_{ab} - \phi_b$.

The blocked field is distributed according to the Boltzmann distribution of another action known as the blocked action, $\phi_b \sim e^{-S_b[\phi_b]}$, where $S_b \equiv B[S_f]$.  Under an arbitrary choice of smoothing kernel $K$, the form of the blocked action can be very different from the lattice action.  We can define an arbitrary lattice action of the form $S(\kappa, \Omega)$, where $\Omega$ represents the infinite space of possible couplings to operators not appearing in \cref{eq:phi4-action} (we ignore $\lambda$ here for simplicity.)  Then the original ``fine action'' $S_f$ describing a field $\phi$ of size $L^2$ can be written as $S_f(\kappa_f, 0)$.  In general, the blocked action
\begin{equation}
S_b = B[S_f(\kappa_f, 0)] = S_b(\kappa_b, \Omega_b)
\end{equation}
includes operators which do not appear in the original lattice action.  In RG language, the fixed point of the block transformation need not lie on the hypersurface $\Omega = 0$.  However, the fixed-point action depends on the blocking, a fact that we can exploit to define an optimized blocking \cite{Hasenfratz:1984hx,Tu:2018dws} that attempts to move the fixed point location as close to the $\Omega = 0$ hypersurface as possible.  If this is achieved exactly, $S_b$ can be identified as a ``coarse action'' $S_c$ which remains on the lattice-action hypersurface:
\begin{equation}
B[S_f(\kappa_f,0)] = S_c(\kappa_{c}, 0).
\end{equation}
This is the ``perfect blocking'' condition.  Because $\kappa$ parametrizes a relevant direction, generally it will flow under RG blocking so that $\kappa_c \neq \kappa_f$; if the system is tuned to a critical point, then $\kappa_c = \kappa_f = \kappa_{\rm cr}$ and perfect blocking in principle leaves the action completely unchanged.

In practice, it is not possible to satisfy the perfect blocking condition exactly, particularly while imposing locality on the smoothing kernel.  Instead, we construct a finite-range smoothing kernel $K$ which acts as  $\psi(x) = \sum_r K(r) \phi(x+r)$ with $|r_i| \leq r_{\rm max}$, and tune its coefficients to minimize the difference between operator expectation values between the blocked and coarse theories,
\begin{equation}
\langle \mathcal{O} \rangle_{B[S_f]} \approx \langle \mathcal{O} \rangle_{S_c}, \label{eq:approx-blocking}
\end{equation}
thereby approximately satisfying the perfect blocking condition.  Optimization is carried out numerically using matched ensembles of field configurations generated at $S_f(\kappa_f,0)$ with volume $(2L)^d$ and at $S_c(\kappa_c,0)$ with volume $L^d$.  Blocking typically alters the correlation length $\xi$ of a system; consequently, it is generally necessary to first tune $\kappa_c$ relative to $\kappa_f$ so that $\xi_c \approx \xi_f/2$.  A correlated $\chi^2$ cost function over several operators (a mix of short-distance and long-distance) is then minimized, comparing the blocked fine ensemble to the coarse ensemble and optimizing the smoothing kernel coefficients.

\paragraph{Upscaling with normalizing flow -- } We impose the invertibility of the smoothing kernel as a constraint during the optimization so that we can readily obtain the original fine field as $\phi = K^{-1} \psi$.  To upscale from a coarse configuration, we must also fill in the detail fields in order to reconstruct the larger $\psi$ fully.  These fields are strongly correlated with $\phi_b$, so we make use of conditional normalizing flow~\cite{Papamakarios:2021,Singha:2026slc} to learn a set of conditional probability distributions for each of the detail fields as
\begin{align}
  d_{01} &\sim q_{\theta,01}(d_{01}\mid \phi_b),\\
  d_{10} &\sim q_{\theta,10}(d_{10}\mid \phi_b,d_{01}),\\
  d_{11} &\sim q_{\theta,11}(d_{11}\mid \phi_b,d_{01},d_{10}).
\end{align}
The composition of these flows defines the inverse blocking map
\begin{equation}
\tilde{\psi}_f = F_\theta(\phi_b, z),\, \, z \sim p_0(z)
\end{equation}
where $p_0(z)$ is a standard Gaussian distribution.  A candidate fine field is obtained by applying the inverse of the perfect smoothing kernel, $\tilde{\phi}_f = K^{-1} \tilde{\psi}_f$.  We use a convolutional neural network architecture for the conditional flows with alternating rational quadratic spline (RQS) \cite{DelDebbio:2021qwf} and affine layers, imposing $\mathbb{Z}_2$ equivariance by making the coupling layers odd in the field $\phi$.  As is standard, we apply flow updates using an even-odd checkerboard so that the Jacobian of the normalizing flow remains triangular and therefore straightforward to evaluate.  To train the network, we take a sample of fine configurations $\{\phi_f\}$ and apply the tuned smoothing kernel to produce $\{\phi_b, d_{ab}\}$ and then minimize the negative-log-likelihood cost function
\begin{equation}
\mathcal{L}_{NLL}(\theta) = -\langle \log {q_\theta(d_{ab}|\phi_b) \rangle}_{\{\phi_b\}}.
\end{equation}

After training, the conditional flow and the inverse smoothing kernel can be used with coarse configurations $\phi_c$ as input instead of $\phi_b$ to obtain proposals for upscaled fine configurations, as  $\psi_{\rm prop} = F_\theta(\phi_c,z)$ and then $\phi_{\rm prop} = K^{-1} \psi_{\rm prop}$.  Due to its convolutional structure, the same flow can be used without adjustment on larger lattice volumes without retraining, although finite volume effects may reduce its effectiveness.

\paragraph{Rethermalization and volume cascade -- }  Instead of sampling from the flow-proposed distribution directly, we treat the upscaled field $\phi_{\rm prop}$ as a proposal and then use local Monte Carlo sweeps targeting the exactly-known fine action $S_f$ to correct its short-distance properties.  After a chosen number of Monte Carlo sweeps, we take the final result $\phi_f$ as a rethermalized fine configuration.

In the limit of a very large number of rethermalization sweeps, $\phi_f$ is guaranteed to be correctly sampled from $e^{-S_f}$. However, the purpose of the initial upscaling step is to correctly seed the long-distance degrees of freedom using physical structure from the matched coarse configuration.  
If this is successful, rethermalization needs only to correct the short-distance degrees of freedom which have short autocorrelation times, requiring only a modest number of local Monte Carlo sweeps.

Once a new fine ensemble $\{\phi_f\}$ with spatial extent $2L$ is obtained from upscaling and rethermalization, we can use that ensemble as the new coarse seed for another upscaling and rethermalization pass.  By iterating this procedure, we can upscale repeatedly from $L \rightarrow 2L \rightarrow 4L \rightarrow 8L ...$, producing a cascaded sequence of larger-volume ensembles.   If the coarse fields $\{\phi_c\}$ used to seed the initial stage are mutually decorrelated, the resulting fine configurations remain independent of each other.  Since each cascade can be evolved independently, the algorithm is also trivially parallelizable along the ensemble index.

\paragraph{Numerical study -- } To demonstrate our algorithm, we carry out numerical studies in the two-dimensional $\phi^4$ theory.  We fix $\lambda=1.0$ and, for simplicity, work  at $\kappa = \kappa_{\rm cr} \simeq 0.340301$ ~ \cite{Bosetti:2015lsa}.  Tuning to criticality allows us to avoid matching coarse and fine ensembles;  we can simply work at fixed $\kappa_{\rm cr}$.  For tuning the perfect smoothing kernel, training the normalizing flow, and also as a basis for comparison, we generate ensembles of 5,000-10,000 independent configurations at lattice volumes from $L=16$ up to $L=128$, using either hybrid Monte Carlo or the Wolff cluster algorithm; the two methods give statistically consistent results.

To tune the smoothing kernel, we take $K(r)$ to be symmetric under the group of lattice rotations $D_4$ and restrict it to a $7 \times 7$ footprint, leaving nine independent coefficients plus the normalization.  We impose the normalization condition $\sum_r K(r) = 2^{\eta/2}$,  with the exact two-dimensional Ising value $\eta=1/4$ \footnote{Alternatively, the normalization can be left floating and determined by the perfect blocking condition as well; doing so in this theory yields values compatible with the exact value, $\eta_\mathrm{pred}\approx 0.251(1)$ with the uncertainty coming from the spread in the optimizer initial conditions.}.  The coefficients of $K(r)$ are optimized against \cref{eq:approx-blocking} by comparing blocked $L=32$ configurations to generated $L=16$ configurations.  Details of the resulting smoothing kernel and the observables used to tune it, as well as additional tests of its effectiveness over observable distributions, are given in the End Matter.

The conditional normalizing flow is trained using the procedure described above on the same $L=32$ fine configurations, using the smoothing kernel optimized in the previous step.  We use the Adam optimizer \cite{kingma2014adam} to minimize the NLL over the network parameters $\theta$.  10\% of the blocked configurations are held out of the training dataset and used for testing.  Further details on the trained conditional flow, including comparisons of upscaled observable distributions to those on directly generated ensembles at $L=64$ and $L=128$, can be found in our longer companion paper \cite{companion}.

Overall, we find that the upscaling procedure closely preserves the distributions of long-distance observables such as the susceptibility $\chi$ or the Binder cumulant $U_4$.  However, distributions of short-distance quantities such as $\langle \phi^2 \rangle$ or the volume-normalized action density $S/V$ slowly degrade as we scale up from smaller to larger volumes.  As a result, our conditional flow does not provide sufficient fidelity to serve as a basis for an independence Metropolis sampler as in \cite{Albergo:2019eim}, but it is well-suited to rethermalization.

For rethermalization, the results shown in our main work use either full-volume or domain-decomposed HMC \cite{Duane:1987de,Luscher:2004pav} with a fixed domain size of $8^2$, and with red-black checkerboarding applied so that domains can be updated in parallel.  Both implementations
give consistent results.  We use HMC deliberately because our method is
intended to generalize beyond scalar models to gauge and fermion systems.   As a cross-check on systematics due to rethermalization, we also study thermalization using the much more efficient Wolff cluster algorithm.

Turning to cascaded upscaling, we find that both the smoothing kernel and the trained normalizing flow generalize well to larger volumes. We successfully ran the upscaling cascade from $L=16$ up to $L=2048$ with no visible degradation in the measured physics results.  The number of rethermalization sweeps required is modest and scales very slowly with the lattice volume; we adopt a fixed number $n^\star = 500$ sweeps at each $L$. Each sweep consists of a trajectory with length $\tau = 2$, with the step size $dt$ tuned so that acceptance rates at each volume remain at approximately 80\%.  This fixed number of sweeps makes the marginal cost of extending an independently seeded cascade to
large volume much smaller than the cost of generating new infrared structure
directly with HMC, whose autocorrelation time for long-distance observables
scales as $\tau\sim L^z$ with $z\approx2.17$ at the two-dimensional Wilson-Fisher
critical point~\cite{nightingale1996dynamic}
\footnote{On the laptop-scale resources used for our numerical studies, extending a
cascade to $L=2048$ requires only tens of seconds, whereas an extrapolation of
direct HMC decorrelation on the same hardware gives a time scale of
order 100 days at the same volume.}.

To assess the validity of the results obtained from our cascaded upscaling procedure, we perform additional analyses to extract physical quantities that can be directly compared with the analytically established values at the two-dimensional Wilson-Fisher critical point. Further measurements and methodological details are provided in a longer companion paper \cite{companion}.

\begin{figure}[t]
\centering
\includegraphics[width=0.485\textwidth]{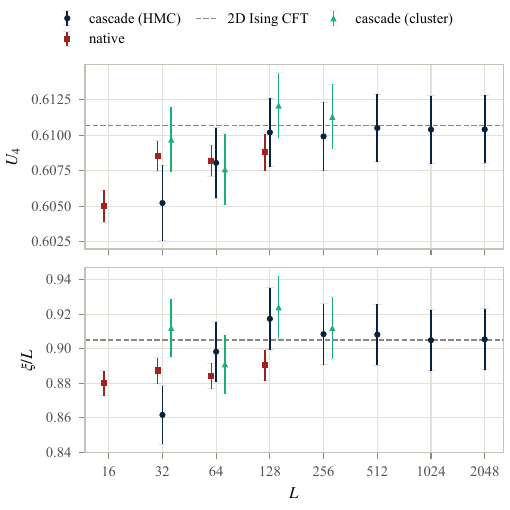}
\caption{Binder cumulant $U_4$ and correlation length $\xi/L$ extracted from the upscaled cascade (black circles) and native HMC/cluster (red squares) ensembles.  Ising CFT results from \cite{Salas:1999qh} are shown as dashed lines for comparison.  Extended rethermalization results using the cluster algorithm (green triangles) are consistent with the HMC cascade.  Cascade results are shown for 2,000 independent chains; native results are for 10,000 configurations.}\label{fig:invariants}
\end{figure}

In \cref{fig:invariants}, we show values of two long-distance RG-invariant observables, the Binder cumulant $U_4$ and the normalized correlation length $\xi/L$, where the latter is computed from the second-moment definition given in \cite{Salas:1999qh,DelDebbio:2021qwf}. The infinite-volume values of these quantities have been calculated \cite{Salas:1999qh} to be $U_4 = 0.6106924(16)$ and $\xi/L = 0.9050488292(4)$.  \Cref{fig:invariants} shows good agreement of our cascade ensembles with both these results and with measurements on native HMC ensembles at smaller $L$. To test for rethermalization systematics, we also show points obtained by extending the cascade ensembles with 50 cluster algorithm sweeps; no evidence of a systematic effect is seen.

An unusual feature of cascade upscaling is that ensembles at different lattice volumes are strongly correlated. For long-distance observables such as $U_4$, the infrared structure is largely inherited from the same small-volume configurations, so combining several volumes in a fit does not provide the statistical gain expected from independent ensembles. 
This correlation is intrinsic to the construction rather than a statistical
defect: the purpose of the cascade is precisely to transport infrared
information already sampled on the root lattice rather than regenerate it at
each larger volume.  Consequently, extending a given cascade to larger $L$ primarily tests the
stability and scale transfer of the method rather than increasing the number
of independent infrared samples.  The inherited infrared structure can also introduce a systematic error if imperfections generated at small $L$ persist through the cascade. Within our present statistical precision, however, we observe no significant evidence for such a bias.

One of the most difficult infrared quantities to evolve in this theory is the signed magnetization, $m=\sum_x\phi(x)$. Because of the $\mathbb{Z}_2$ symmetry under $\phi\rightarrow-\phi$, its distribution has two peaks, and tunneling between them becomes increasingly suppressed at large volume, even for highly efficient sampling algorithms such as Wolff cluster. In cascade upscaling, the relative population of the two $\mathbb{Z}_2$ sectors is inherited from the small-volume ensemble that seeds the cascade, preserving good sampling of both peaks up to our largest volume, $L=2048$. More generally, even when substantial rethermalization is required, cascade upscaling can provide large-volume initial conditions that inherit the sampling of slowly evolving sectors from a smaller lattice. This suggests a possible application to systems with severe sector freezing, such as topological sectors in lattice gauge theory.

\paragraph{Conclusion -- } We have introduced a novel inverse-RG algorithm for ``cascade upscaling'', which generates large-volume ensembles with efficient sampling of long-distance modes.  In the two-dimensional $\phi^4$ theory at the Wilson-Fisher critical point, we have demonstrated upscaling from $L=16$ to $L=2048$ with no observable degradation of the physical observables tested.  Our algorithm combines approximate perfect blocking, conditional reconstruction of the missing detail fields, and short rethermalization with the target action.  It  mitigates critical slowing down by taking independent small-volume configurations to large-volume ones with the correct physics.  The inverse blocking must be corrected by local rethermalization sweeps, but since only short-distance modes need significant repair, the cost scaling with volume is extremely modest.  Our algorithm is also trivially parallelizable over the size of the ensemble, since each cascaded upscaling of an initial small-volume configuration is independent of the others.

Although we have worked at the Wilson-Fisher critical point in scalar field theory for this demonstration, we do not believe there are any major obstacles to generalizing the proposed method to gauge or fermion degrees of freedom.  In addition, the method should be applicable away from criticality; indeed, any potential systematics associated with the rethermalization step may be easier to control in systems with finite correlation length.  

\paragraph{Acknowledgments --}
We thank Dan Hackett, Jake Sitison, and Roman Marcarelli for useful discussions.  This research was partially supported by DOE grant DE-SC0010005.  Code development was performed with substantial assistance from OpenAI's ChatGPT and Codex and Anthropic's Claude Opus 5 and Fable 5.  A.H. and E.T.N. directed the development and carried out the debugging, validation, and testing of the final implementation. A.H. acknowledges the stimulating environment at ECT* Trento during the June 2023 workshop ``Machine Learning for Lattice Field Theory and Beyond'', which  inspired her work on inverse RG and normalizing flows.

\bibliographystyle{apsrev4-2}
\bibliography{rg_guided_nf_refs}

\appendix

\section{End Matter}

\paragraph{Operator definitions --} For training, testing, and validation of the various components of our algorithm, we measure a variety of standard observables which we briefly define here.  For short-distance probes of the theory, we measure the ultralocal operators
\begin{equation}
\phi^{2n} \equiv \frac{1}{V} \sum_x \phi_x^{2n}
\end{equation}
and the $D_4$-symmetric two-point correlation function
\begin{equation}
G(r) \equiv \frac{1}{\mathcal{N}_r} \sum_{r' \in \mathcal{O}_r} \frac{1}{V} \sum_x \phi_x \phi_{x+r'}
\end{equation}
where $\mathcal{O}_r$ represents the $D_4$ symmetry orbit associated with $r$ and $\mathcal{N}_r$ is its multiplicity.  In two dimensions we write these as $G_{ij}$ where $i \geq j$ are the components of the vector $r$ defining the orbit; for example, $G_{10}$ is the nearest-neighbor bilinear which appears in the action.

To probe long-distance structure, we consider the magnetization
\begin{equation}
m \equiv \frac{1}{V} \sum_x \phi_x
\end{equation}
and its moments; due to $\mathbb{Z}_2$ symmetry $\langle{m \rangle} = 0$ is expected, but higher moments like $\langle m^2 \rangle$ are non-vanishing.  We also consider the momentum-space correlation function
\begin{equation}
G(p) \equiv \frac{1}{V} \left| \sum_x e^{ip \cdot x} \phi_x \right|^2,
\end{equation}
and $G(p_{\rm min})$ which is evaluated at the lowest non-zero lattice momentum $|p_{\rm min}| = 2\pi/L$, averaged over the two directions.  Two key derived quantities are the Binder cumulant $U_4$ and the second-moment estimator of the correlation length $\xi$, given by
\begin{equation}
U_4 \equiv 1 - \frac{\langle m^4 \rangle}{3\langle m^2 \rangle^2},\; \; \xi = \frac{1}{2\sin (\pi/L)} \sqrt{\frac{\chi}{G(p_{\rm min})} - 1},
\end{equation}
where the susceptibility $\chi \equiv V \langle m^2 \rangle$.

\begin{figure}[t]
      \centering
      \includegraphics[width=0.78\linewidth]{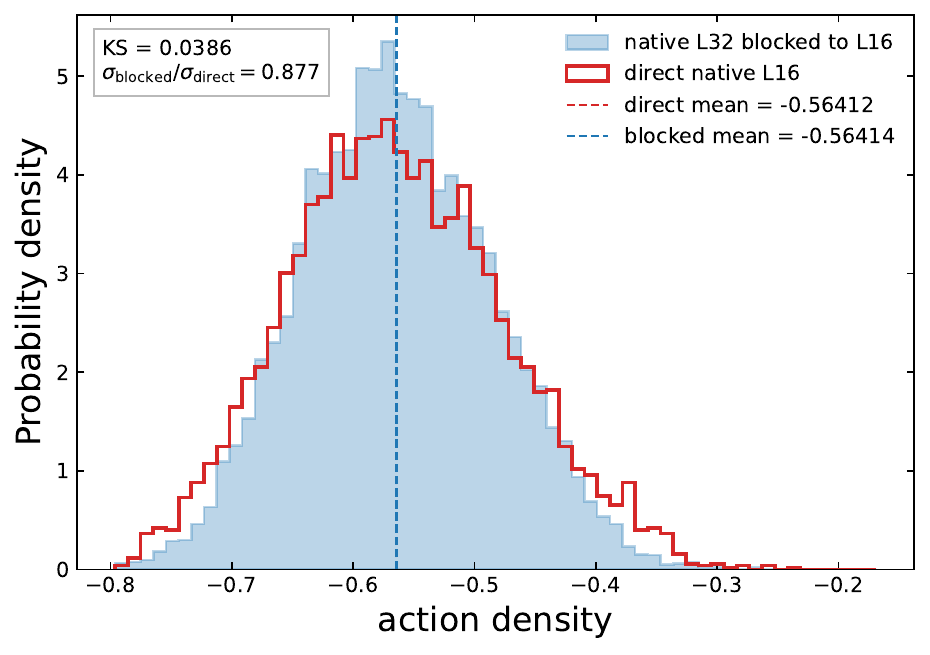} \\
      \includegraphics[width=0.78\linewidth]{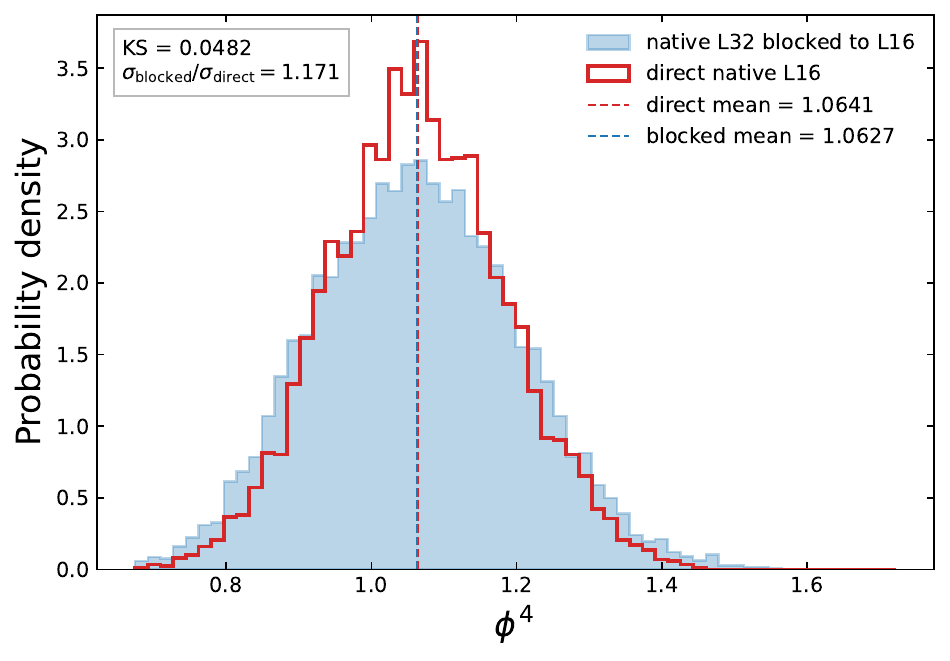} \\
      \includegraphics[width=0.78\linewidth]{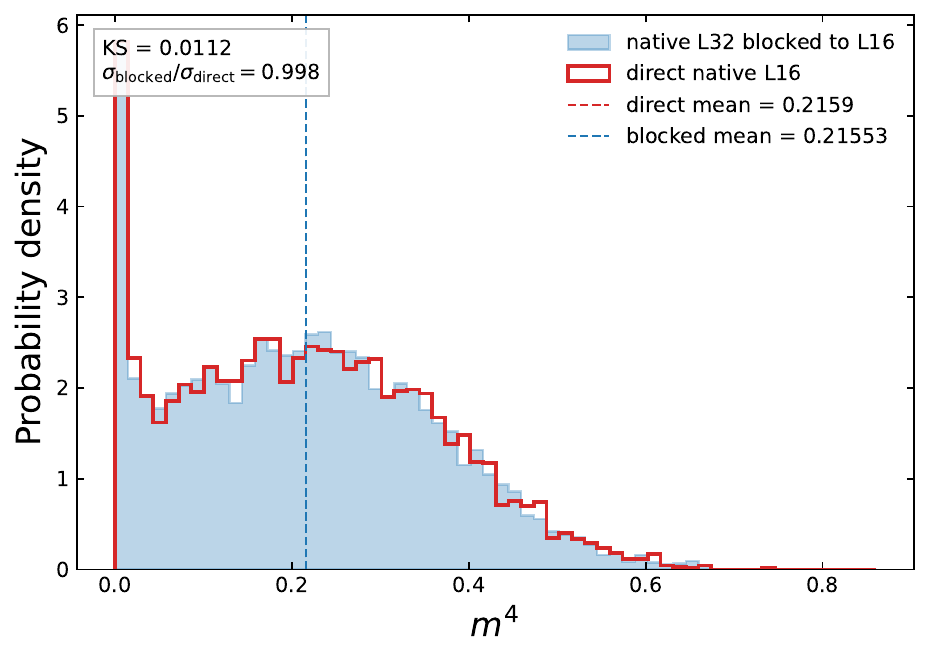}
      \caption{Approximate perfect blocking test for $L32\to L16$ blocking.  Direct coarse configurations and blocked fine configurations should have overlapping  distributions. The panels compare the action density and $\langle \phi^4\rangle$ distributions, and the long distance quantity $m^4$. }
      \label{fig:L32-to-16-blocking-histograms}
\end{figure}

'\begin{figure}[b]
      \centering
      \includegraphics[width=0.78\linewidth]{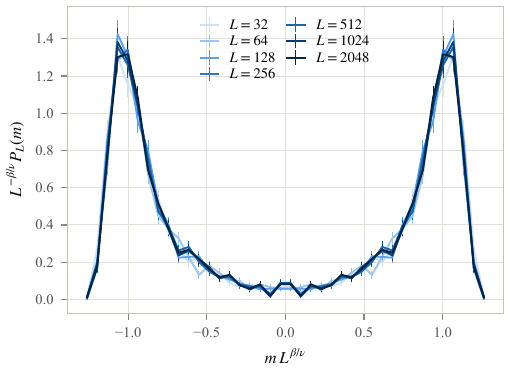}
      \caption{Distribution of the magnetization $m$ at each volume in the cascade ensemble, showing the expected double-peaked structure and scaling with the volume according to the exact critical exponents.}
      \label{fig:m-histogram}
\end{figure}

\paragraph{Smoothing kernel --}
Details of the definition for the smoothing kernel $K$ are given in the main text.  It is optimized so that the distribution of $\psi_{\ee}\equiv D_{ee}\,K\,\phi$ obtained by blocking fine configurations matches the distribution of directly generated coarse configurations.

The fitting objective minimizes a covariance-weighted mismatch of various observables,
  \begin{equation}
  \chi^2_K = \sum_{a,b}\left(\langle O_a\rangle_{B\Sf}
  -\langle O_a\rangle_{\Sc}\right)
  C^{-1}_{ab}
  \left(\langle O_b\rangle_{B\Sf}
  -\langle O_b\rangle_{\Sc}\right).
  \label{eq:kernel-objective}
\end{equation}

  The fitted operator set is
  \begin{equation}
   {\cal O}_{\rm fit}=
   \left\{
   \phi^2,\phi^4,\phi^6,G_{10},G_{20},G_{11},
   G_{21},G_{22},G_{30},G_{31},m^2
   \right\}.
  \end{equation}

  The operators $m^4$, the local kurtosis
  $\langle\phi^4\rangle/\langle\phi^2\rangle^2$, the action density, and
  $G(p_{\min})$ are reserved for validation rather than included in the
  fit.  We impose that the kernel obtained must be invertible, i.e. $K(p)$ does not have any zeros.  The kernel we obtain based on tuning $L=32\to16$ is shown in Table \ref{tab:phi4_kernel_orbits}.

 \begin{table}[tbh]
  \centering
  \caption{
    Optimized $7\times7$ $\phi^4$ blocking kernel, represented by
    $D_4$ symmetry orbits about the central site. The tabulated coefficients
    already include the anomalous-dimension normalization factor
    $2^{\eta/2}$; no additional factor is applied.
  }
  \label{tab:phi4_kernel_orbits}
  \begin{ruledtabular}
  \begin{tabular}{ccc}
  Offset orbit $(|\Delta x|,|\Delta y|)$ & Multiplicity & $K(\Delta x,\Delta y)$ \\
  \hline
  $(0,0)$ & 1 & $ 0.888641822$ \\
  $(1,0)$ & 4 & $ 0.010508374$ \\
  $(1,1)$ & 4 & $-0.066254410$ \\
  $(2,0)$ & 4 & $ 0.037116968$ \\
  $(2,1)$ & 8 & $ 0.022657096$ \\
  $(2,2)$ & 4 & $-0.001649881$ \\
  $(3,0)$ & 4 & $ 0.020222068$ \\
  $(3,1)$ & 8 & $ 0.007299892$ \\
  $(3,2)$ & 8 & $-0.003848342$ \\
  $(3,3)$ & 4 & $-0.001693935$ \\
  \end{tabular}
  \end{ruledtabular}
  \end{table}

The conditioning factor of the kernel is defined in Fourier space
\begin{equation}
  \kappa_K=\frac{\max_{\mathbf p}|K(\mathbf p)|}
  {\min_{\mathbf p}|K(\mathbf p)|}.
\end{equation}
For the selected kernel, $\min_{\mathbf p} K(\mathbf p)=0.54372$ and
$\max_{\mathbf p} K(\mathbf p)=1.24390$, so $\kappa_K=2.2878$ and
$\max_{\mathbf p}|K(\mathbf p)^{-1}|=1.8391$.
  The nonzero minimum is the key invertibility check; the modest condition number means inversion does not strongly amplify modes.

An essential diagnostic is distributional overlap.  Fig.~\ref{fig:L32-to-16-blocking-histograms} compares the histograms of blocked fine fields and direct coarse fields of the action-density, $\langle \phi^4\rangle$, and the long-distance $m^4$ (which was held out of the training) distributions for $L=32 \to L=16$.  To quantify the agreement, the plots give the ratio of the widths of direct versus blocked distributions
$R_\sigma = \sigma_{\rm blocked}\,/\,\sigma_{\rm direct}$ and
the Kolmogorov-Smirnov (KS) distance between the two  cumulative distributions.  Similar tests on larger volumes $L=64 \to L=32$, and $L=128 \to L=64$ maintain good properties for long-distance quantities and continue to reproduce short-distance distributions well; see the companion paper \cite{companion} for more plots and tests.

\paragraph{Magnetization distribution --}
Finally, Fig.~\ref{fig:m-histogram} shows the histogram of the magnetization $m$ over our cascade ensembles, rescaled with $L^{\beta/\nu} = L^{1/8}$.  
The rescaled distributions retain the expected
double-peaked structure, with good sampling of both peaks through the largest volumes in the cascade.

\end{document}